\documentclass[pra,12pt,tightenlines,notitlepage]{revtex4-1} 
\usepackage{amssymb,amsmath,amsfonts,bm,bbold,amssymb,amsbsy,latexsym,amsthm,dsfont,array,layout,mathrsfs,mathdots,ucs,centernot,graphicx,bbm}
\usepackage[normalem]{ulem}
\usepackage[usenames,dvipsnames]{color}
\usepackage[usenames,dvipsnames,svgnames,table]{xcolor}
\usepackage{tikz}
\usetikzlibrary{arrows,fit,positioning,automata}
\usetikzlibrary{decorations,decorations.pathmorphing,decorations.pathreplacing,shapes.arrows,shapes.geometric}

\newcommand\twoheaduparrow{\mathrel{\rotatebox[origin=c]{90}{$\twoheadrightarrow$}}}
\newcommand\twoheaddownarrow{\mathrel{\rotatebox[origin=c]{-90}{$\twoheadrightarrow$}}}
\newcommand{\ket}[1]{\left| #1 \right>}

\newcounter{thm}
\newcounter{co}
\newcounter{re}
\newtheorem{Proposition}[thm]{Proposition}
\newtheorem{Remark}[re]{Remark}
\newtheorem{Condition}[co]{Condition}
\newtheorem{Definition}[thm]{Definition}

\begin{document}

\title{Quantifying Contextuality}

\author{Ian T. Durham }
\email[]{idurham@anselm.edu}
\affiliation{Department of Physics, Saint Anselm College, 100 Saint Anselm Drive, Manchester, NH 03102, United States}

\begin{abstract}
In this essay, I develop order-theoretic notions of determinism and contextuality on domains and topoi. In the process, I develop a method for quantifying contextuality and show that the order-theoretic sense of contextuality is analogous to the sense embodied in the topos-theoretic statement of the Kochen--Specker theorem. Additionally, I argue that this leads to a relation between the entropy associated with measurements on quantum systems and the second law of thermodynamics. The idea that the second law has its origin in the ordering of quantum states and processes dates to at least 1958 and possibly earlier. The suggestion that the mechanism behind this relation is contextuality, is made here for the first time.
\end{abstract}


\maketitle

\section{Background}\label{sec1}

The theories of categories, topoi and domains have all been extensively applied to the study of physical systems, both quantum and classical~\cite{Martin:2000fk,Knuth:2003vn,Coecke:2006aa,Abramsky:2008uq,Isham:2010kx,Spivak:2013fk}. In this essay, I extend the work of Coecke and Martin~\cite{Coecke:2011uq,Martin:2011fk} on states and measurements, within the neo-realist framework developed by D\"{o}ring and Isham~\cite{Doring:2011fk}, to the concept of quantum contextuality. I begin with a review of the basic principles and definitions that will be used throughout this essay. 

A category is a mathematical structure that consists of objects and ``arrows'', along with the requirements of compositeness, associativity and identity. The arrows (called morphisms) represent a connective pattern between the objects. A category is anything that satisfies these conditions~\cite{Awodey:2010uq,Spivak:2013fk}. Consider a physical system $S$. In general, $S$ can be described by real-valued physical quantities (or expectation values) $\mathcal{Q}$ assigned by the state of the system $\rho$. In other words, there is a distinction between a physical quantity and its real-valued representation. For example, consider an ideal gas whose pressure is measured to be 10 Pa. Pressure is the physical quantity, whereas ``10'' is the real-valued representation of that quantity at a given instant. The state of the system $\rho$ represents a specific mapping from the physical quantities of the system to the real-valued representations of those quantities, $\rho : \mathcal{Q} \to \mathbbm{R}$. 

This description has two notable features. The first is that in proposing this description, D\"{o}ring and Isham adopt a specifically ``neo-realist'' structure. They interpret $\mathcal{Q}$ and $\mathbbm{R}$ as objects in a category, referring to them as the state-object and the quantity-value object, respectively. As such, the category of which they are a part is of a special type known as a topos. As they note, ``[w]hatever meaning can be ascribed to the concept of the ``value'' of a physical quantity is encoded in (or derived from) this representation''~\cite{Doring:2011fk}. In this manner, their interpretation is technically independent of any type of agent, \textit{i.e.}, no agent is necessary to instantiate the state. This is intentional on their part for reasons related to quantum gravity. However, it should be noted that their description does not strictly preclude the existence of an agent. The framework is generally agnostic on the issue and allows for a variety of interpretations, both with and without agents. 

The second notable feature of this model is that it lends itself to a description that encompasses both classical and quantum systems under a single structure. Thus, $\rho$ can represent the state of a classical, quantum, or hybrid system, as long as it is more generally viewed as an arrow in a topos. One final cautionary note should be made regarding topoi, however. The aforementioned description focuses on one of three primary requirements for defining topoi. The other two are quite technical, and I will only briefly mention them. One is that propositions about a system are represented by sub-objects of the state-object. These sub-objects form a Heyting algebra (as do any sub-objects of any object within the topos). The other is that, in general, an object in a topos may not be determined by its ``points'' \textit{per se}. As such, microstates are not always useful (though they are not strictly prohibited). In fact, in quantum theory the state object has no points (microstates) at all. D\"{o}ring and Isham show that this lack of microstates is equivalent to the Kochen--Specker theorem.

In this manner, D\"{o}ring and Isham confront the basic (and age-old) question, ``What is a thing?'' Their definition bears a strong resemblance to Eddington's conception of a physical ``object'' as a somewhat loosely defined bundle of properties wherein these properties refer to values of physical quantities~\cite{Eddington:1939fk}. In other words, an object is defined by its state $\rho$. This means that for the purposes of this paper, we take an object or system to be equivalent to its state, \textit{i.e.}, $S:=\rho$. Even though the state itself is a mapping, it is imperative to remember that this does not necessarily require the existence of an agent. Indeed, D\"{o}ring and Isham are adopting an expressly neo-realist view. Nevertheless, neither does it preclude such an agent. This is a very subtle point that is somewhat counter-intuitive as we are used to thinking of states as being representations of physical states. Within the present framework, however, we have essentially abstracted away the difference. Thus, $\rho$ can be both an object and an arrow in a category (and, indeed, category theory itself recognizes this duality: a slice category, for example, has arrows as its objects~\cite{Awodey:2010uq}). In the case of a complex physical system $\mathcal{S}$ that is composed of smaller physical systems $S \in \mathcal{S}$, we generalize this by saying that the state of $\mathcal{S}$ would be given by the set $\Sigma$ of all sub-system states $\rho_i \in \Sigma$, where $\rho_i : \mathcal{Q}_i \to \mathbbm{R}_i$. 

Let us now consider a set of such sub-system state-objects $\Sigma$ together with a partial order $\sqsubseteq$ that includes certain intrinsic notions of completeness and approximation that are defined by this order. Together, they form a domain, $(\Sigma,\sqsubseteq)$. Given two objects $\rho , \sigma \in \Sigma$, the statement $\rho \sqsubseteq \sigma$ is interpreted as saying that $\rho$ contains (or carries) some information about $\sigma$, \textit{i.e.}, $\sigma$ is ``more informative'' than $\rho$~\cite{Martin:2011fk}. We take ``information'' to be anything that may be represented by a state $\rho$. In the event that $\rho$ contains complete information about $\sigma$, then $\rho = \sigma$ and $\rho$ is said to be a maximal element (object) of the domain, in which case it is an example of an ideal element. An object that is not ideal is said to be partial. The order $\sqsubseteq$ is interpreted as an information order in the sense that if a process generates some ordered sequence $(\rho_{n})$ of elements that increases with respect to the information order, then $\rho_{n}\sqsubseteq \rho_{n+1}$ for all $n$, \textit{i.e.}, $\rho_{n+1}$ is more informative than $\rho_{n}$. To be clear, information is defined as anything that can be represented by $\rho$; no ordering relation is necessary. As such, not all information necessarily has an inherent order. A special type of information order is an approximation $\preceq$, where the statement $\rho \preceq \sigma$ is interpreted as saying that $\rho$ approximates $\sigma$~\cite{endnote1}. We take this to mean that $\rho$ carries essential information about $\sigma$. In other words, any ``information path'' leading to $\sigma$ must pass through $\rho$. An element $\rho$ of a poset is compact if $\rho \preceq \rho$.

If one takes a neo-realist viewpoint, information about reality exists independent of any agent or observer. On the other hand, it often only makes sense to talk about ordering information in the context of an agent or observer who is obtaining that information. The ordering relation $\sqsubseteq$ has to do with potential evolutions of a state and depends directly on the agent, whereas $\preceq$ has to do with the intrinsic information about a state. For example, consider a set of three closed boxes. We are told that there is a ball in one of the three boxes, $A$, $B$ and $C$. Our task is to determine the color of the ball. Clearly, in order to determine its color, we must first locate it. Suppose we open box $A$ and find that it does not contain the ball. We only have two boxes left; we have eliminated one possibility. Thus, the states of the other two boxes are clearly more informative than the state of the box we just opened, \textit{i.e.}, $\rho_A \sqsubseteq \rho_B, \rho_C$. If we open box $B$ and do not find the ball, we immediately know it is in box $C$. Thus, $\rho_A \sqsubseteq \rho_B \preceq \rho_C$. We may now observe the ball to determine its color. Suppose, instead, that we opened box $B$ first. In that case, $\rho_B \sqsubseteq \rho_A \preceq \rho_C$. The neo-realist view assumes that the presence of the ball in box $C$ is independent of any agent: $\rho_C$ contains essential information about the ball's color, regardless of whether or not the box is opened. Thus, the relation $\preceq$ is independent of any agent, whereas $\sqsubseteq$ is not.

Both ordering relations can be thought of as a collection of arrows where each arrow points from one state (mapping) $\rho_i$ to another. As such, the domain consisting of the objects $\rho_i$ along with the set of arrows associated with the information order form a category. Each object $\rho_i$ by itself is, of course, an arrow $\rho_i : \mathcal{Q}_i \to \mathbbm{R}_i$ in a topos with a state-object $\mathcal{Q}_i$ and a quantity-value object $\mathbbm{R}_i$. Each of the objects $\rho_i$ represents some amount of partial information about the larger system, as noted above. It is the amount of partiality that helps to define the order. In order to quantify this, we now more formally define a measurement $\mu$ on a domain $\Sigma$ as a function $\mu : \Sigma \to \mathbbm{R}^{+}$ that assigns to each informative object on the domain, $\rho_i \in \Sigma$, a number $\mu\rho_i \in \mathbbm{R}^{+}$ that measures the amount of partiality contained in~$\rho_i$. This number is referred to as the information content of the object $\rho_i$. Thus, moving ``up'' a given information order (such as the order in which we opened the boxes in search of the ball) represents decreasing partiality,~\textit{i.e.},
\begin{equation}
\rho \sqsubseteq \sigma \; \Rightarrow \; \mu \rho \ge\mu \sigma.
\label{eqinfo}
\end{equation}
Further, if $\rho$ is a maximal element of a domain, then $\mu\rho = 0$~\cite{Coecke:2011uq}. Such a state is said to be pure. Intuitively, this means that if full knowledge of an object is known, then there is obviously no partiality. It is important to remember that $\mu$ may be any number as long as~\eqref{eqinfo} and the condition:
\begin{equation}
\mu\rho=0 \Rightarrow \rho\in\max(\Sigma)
\label{eqcond}
\end{equation}
are satisfied. This requirement then constrains the possible functional forms that $\mu$ may take. There are several conditions that are implicit in~\eqref{eqinfo} and~\eqref{eqcond}. The key is recognizing that we are quantifying an ordering relation of mappings based on how informative each mapping is relative to the others, \textit{i.e.}, we seek to essentially rank these mappings. For example,~\eqref{eqinfo} tells us that $\sigma$ is more informative than $\rho$ and, thus, essentially ranks $\rho$ and $\sigma$ according to the information that they provide.

Consider a measurement $\mu\rho$ on a state $\rho$. The measurement will presumably have a number of possible outcomes. Whatever those outcomes may be, the information supplied by the measurement provides a certain amount of insight into the object or system under consideration. Suppose we now add additional outcomes, all of which are zero probability. It seems trivial to assume that adding these zero-probability outcomes will not add or subtract from the information provided by a measurement, \textit{i.e.}, that the measurement should not supply us with any more or any less information than it would without the additional zero-probability outcomes. This assumption is known as expansibility, and it implies that adding zero-probability outcomes to a given measurement will not change the ordering relation~\eqref{eqinfo}. In the example involving the boxes and the ball, this would be akin to adding a third outcome that has zero likelihood of occurring, say, for example, the possibility that the ball is both in the box and not in the box simultaneously (which, for classical systems, such as balls and boxes, is clearly absurd). The addition of this outcome does not fundamentally alter the ordering relation of the states of the various boxes. (Note that this is especially true in a neo-realist interpretation.)

It is also natural for us to assume that there is symmetry in the outcomes of any given measurement, \textit{i.e.}, permuting them does not change the amount of information given by the measurement. In other words, it is irrelevant as to whether or not we find the ball in a given box; the presence or absence of the ball are of equal ``worth'' to us as far as an individual measurement is concerned.

Now, consider two measurements $\mu\rho$ and $\mu\sigma$ on a pair of states $\rho$ and $\sigma$, respectively. We assume that the information gained from a joint measurement on the two systems is less than or equal to the sum of the information gained from individual measurements on the two systems. This, of course, is the well-known property of subadditivity ({\em cf.}~\cite{Lieb:1975aa,Nielsen:2000aa}). When the information gained from the joint measurement is identical to the sum of the information from the individual measurements (\textit{i.e.}, when equality holds), the information is said to be additive~\cite{Aczel:1974aa}. While subadditivity appears intuitive, there strangely do exist systems that may be additive, but not subadditive~\cite{endnote2}.

Thus, an appropriate functional form for $\mu$ would satisfy both~\eqref{eqinfo} and~\eqref{eqcond}, as well as the conditions of expansibility, symmetry, subadditivity and additivity (as well as a standard normalization condition). Acz\'{e}l, Forte and Ng have shown that only linear combinations of the Shannon and Hartley entropies satisfy the conditions of expansibility, symmetry, subadditivity, additivity and normalization. Coecke and Martin have shown that Shannon entropy satisfies~\eqref{eqinfo} and~\eqref{eqcond}. Thus, we are justified in choosing the Shannon entropy:
\begin{equation}
\mu\rho =-\sum_{i=1}^{n}\rho_{i}\log \rho_{i}
\label{eqent}
\end{equation}
as a functional form of $\mu$, where the individual mappings $\rho_i$ must be chosen, such that they yield real-number representations that allow~\eqref{eqent} to satisfy~\eqref{eqinfo} and~\eqref{eqcond}~\cite{endnote3}. Note also that Acz\'{e}l {\em et al.} employ Forte's interpretation of ``experiments (measurements)'' and ``outcomes'' as partitions of a set~\cite{Forte:1975aa}. This is particularly appropriate for our purposes, given that the domain-, category- and topos-theoretic approach presented here is fundamentally built on a set-theoretic foundation. Note also that Forte shows that Shannon entropy is the only function defined on $n$-tuples that fully satisfies the conditions of expansibility, symmetry, subadditivity and additivity (as well as of probabilistic normalization)~\cite{Forte:1975aa}. Thus, though the Shannon entropy is technically a scalar, we are interested in its ability to rank the mappings, which allows us to preserve the structure of the $n$-tuple in the ordering relation itself. 

A decreasing partiality, then, corresponds to decreasing uncertainty about a physical system, such that when the complete state is known, there is no uncertainty about it. As such, regardless of the functional form that $\mu$ takes, we use the term ``relative'' entropy to describe anything that satisfies~\eqref{eqinfo} and~\eqref{eqcond}. Notice that this is entirely consistent with the usual counter-arguments to Maxwell's Demon, since it necessarily requires an agent (hence, the use of the term ``relative''). One might assume that in adopting the general domain-theoretic definition of a measurement as a broader definition for entropy, I am taking entropy to be a measure of knowledge where, in this sense, ``knowledge'' refers to information transferred to an agent. Thus ``knowledge'' would necessarily be an agent-dependent concept. In fact, it is the exact combination of an agent and a neo-realist view that raises the false specter of Maxwell's Demon in the first place: the only reason the entropy decreases to begin with is precisely because of the presence of the agent (``demon''), whose own entropy is not being properly accounted for. To be clear, I am expressly not adopting this view of entropy. Rather, entropy is interpreted here solely as a method of rank-ordering states based on their relative informativeness about a system.

Returning to the mathematical aspects of measurements on domains, consider some information order on a domain, such that $\rho_1 \sqsubseteq \rho_2 \sqsubseteq \rho_3 \ldots \preceq \rho_n$. The sense of decreasing partiality suggests that the information order is, in Martin's words, ``going somewhere''~\cite{Martin:2011fk}. That is:
\begin{equation}
\rho_1 \sqsubseteq \rho_2 \sqsubseteq \rho_3 \ldots \Longrightarrow \bigsqcup_{n\in\mathbbm{N}}\rho_n \in \Sigma
\label{eq:sup}
\end{equation}
where the element $\bigsqcup_{n\in\mathbbm{N}}\rho_n$ is a maximal element of the domain. Thus, any process that leads to \linebreak $\rho = \bigsqcup\rho_n$ will yield an entropy of:
\begin{equation}
\mu\left(\bigsqcup_{n\in\mathbbm{N}}\rho_n\right)=\lim_{n\to\infty}\mu\rho_n.
\label{eq:supmeas}
\end{equation}

With this in mind, it makes sense to ask if there is some way in which an agent can maximize the efficiency associated with obtaining the information about a particular domain. For instance, it might be that under one particular choice of information order, two of the chosen sub-states contain redundant information. Ideally, any set of sub-states considered by an agent should be maximally informative with no redundancy, \textit{i.e.}, no two sub-states should contain duplicate information, but all of the sub-states considered together should give complete information about the state. To better understand this point, consider a simple system that may be completely characterized by a single measurement (``query'') that yields a ``yes'' or a ``no'' to the query. The state of the system is thus a map $\rho : \mathcal{Q} \to \{0,1\}$, and the purpose of a measurement by an agent is to distinguish between $\rho : \mathcal{Q} \to 0$ and $\rho : \mathcal{Q} \to 1$, \textit{i.e.}, to determine which of the two possible states the system is actually in. Maximizing the information requires choosing the correct basis within which to measure the system. In fact, Schumacher and Westmoreland define information as the probability of successfully distinguishing between orthogonal measurements~\cite{Schumacher:2010uq}.

These ideas are nicely summarized in a mathematical form via a directed-complete partial order or dcpo. Intuitively, a dcpo is a poset in which every directed sub-set has a supremum. In other words, every sub-state of a state should be maximally knowable, by which I mean that the maximum amount of information for a sub-state should be, under ideal conditions, fully transferable to an agent if one exists. Formally, a dcpo is defined as follows.
\begin{Definition}[dcpo]
Let $(\mathcal{P},\sqsubseteq)$ be a poset. Then, the nonempty subset $P \subseteq \mathcal{P}$ is said to be directed if for all $x,y\in P$ there exists $z \in P$, such that $x,y\sqsubseteq z$. The supremum $\bigsqcup P$ of $P\subseteq \mathcal{P}$ is the least of its upper bounds when it exists. A dcpo is a poset in which every directed set has a supremum.
\end{Definition} 
Any continuous dcpo is an example of a domain~\cite{Martin:2011fk,Coecke:2011uq}. 

In the example in which a colored ball was in one of three boxes, we considered a sequence of processes that resulted in one of two possible outcomes in each case: either the ball was in the box or it was not. Any time we have a situation in which a process has more than one possible outcome (be it two or ten), we need a formal way to distinguish between those outcomes. Consider, then, $n+1$ boxes and assume that within one of these boxes is a ball~\cite{endnote4}. Now, suppose that Alice and Bob are each tasked with locating the ball and determining its color. Prior to opening any boxes, the state of the system, representing both Alice's and Bob's knowledge, is given by the completely mixed state:
\begin{equation}
\rho_{\perp} = (1/(n+1),\ldots,1/(n+1))
\label{eq:mixed}
\end{equation}
where we are representing the fact that Alice and Bob both initially assume that the ball is equally likely to be in any of the boxes. In this case, the state (or their knowledge of the state, if you prefer) is a probability. Once the ball is located, the state is then given by the pure state:
\begin{equation}
\rho_{i} = (0,\ldots,1,\ldots,0)
\label{eq:pure}
\end{equation}
where $i$ indicates in which of the $n+1$ boxes the ball is found. In a neo-realist view,~\eqref{eq:pure} is independent of any agent. If we say that $\rho \in \Sigma_{n+1}$ represents the state as it appears to Alice and $\sigma \in \Sigma_{n+1}$ represents the state as it appears to Bob, then the statement $\rho\sqsubseteq\sigma$ indicates that Bob knows more about the position of the ball than Alice. For example, perhaps Bob was able to look in the boxes faster than Alice. For every box in which the ball is not found, the state can be updated, since one possibility has been eliminated. As such, Bob could eliminate possibilities faster than Alice. In this way the completely mixed state is the least element of the domain of states $\Sigma_{n+1}$. The set of all pure states would thus be the set of all maximal elements. Coecke and Martin use a similar example to show that there exists a unique order on classical two-states given by the set $\Sigma_2$ and that a partial order $\sqsubseteq$ on the more general $\Sigma_n$ respects a mixing law under certain restrictions~\cite{Coecke:2011uq}. The most important point here is that classical states have a unique ordering relation. I will return to this later.

Now, let $(\Sigma,\sqsubseteq)$ be a dcpo. For elements $\rho,\sigma \in \Sigma$, we write $\rho \preceq \sigma$ if and only if for every directed subset $\Delta\subseteq\Sigma$ with $\sigma\sqsubseteq\bigsqcup \Delta$ we have $\rho\sqsubseteq \tau$ for some $\tau\in \Delta$. In order to simplify notation, we introduce the following sets:
\begin{equation}
\begin{aligned}
\twoheaduparrow \rho := \{\rho,\sigma\in \Sigma: \rho\preceq \sigma\} && \textrm{and} & & \twoheaddownarrow \rho := \{\rho,\sigma\in \Sigma : \sigma \preceq \rho\} \\
\uparrow \rho := \{\rho,\sigma\in \Sigma: \rho\sqsubseteq \sigma\} & & \textrm{and} & & \downarrow \rho := \{\rho,\sigma\in \Sigma: \sigma\sqsubseteq \rho\} \label{eq:direct}
\end{aligned}
\end{equation}
where the arrows suggest the ``direction'' of the information order. In fact, these may be viewed as defining a specific information order. Therefore, for example, $\uparrow\rho$ is the set of states for which $\sigma$ is more informative than $\rho$, whereas $\uparrow\sigma=\;\downarrow\rho$ is the set of states for which $\rho$ is more informative than $\sigma$. Thus, for some dcpo $(\Sigma,\sqsubseteq)$, a pair of elements $\rho,\sigma \in \Sigma$ are said to be orthogonal if:
\begin{equation}
\mu(\uparrow \rho\;\cap\uparrow \sigma)\in \{0\}
\label{eq:ortho}
\end{equation}
where $\{0\}$ is the null set and $\Sigma$ is assumed to have a least element $\rho_{\perp}$. In a way, this formalizes the neo-realist viewpoint, since it says that there can be no reality in which $\sigma$ is more informative than $\rho$ and simultaneously $\rho$ is more informative than $\sigma$, \textit{i.e.}, there exists only one reality for a set of processes. This offers another way to distinguish between the relations $\sqsubseteq$ and $\preceq$: the former is a statement about knowledge and, hence, as mentioned above, is agent-dependent, whereas the latter is a statement about processes and (in a neo-realist interpretation) is agent-independent. Crucially, this is related to the fact that, as Coecke and Martin have shown, there exists a unique order on classical states~\cite{Coecke:2011uq}. I discuss this further in Section~\ref{sec2}.

Now consider a specific agent's lack of information. We can quantify this lack of information via the Shannon entropy (Equation~\eqref{eqent}), such that conditions~\eqref{eqinfo} and~\eqref{eqcond} hold. In other words, for some dcpo ($\Sigma,\sqsubseteq$) and $\rho,\sigma\in\Sigma$, we set:
\begin{equation}
\mu \rho =-\sum_{i=1}^{n}\rho_{i}\log \rho_{i} \quad \textrm{with} \quad \rho \sqsubseteq \sigma \;\Rightarrow\; \mu \rho \ge \mu \sigma
\label{eq:entcond}
\end{equation}
where for some value $n=N$, we have $\rho=\sigma$ and $\mu \rho = \mu \sigma$. Thus, as our knowledge of $\Sigma$ increases, $\rho$ approaches the maximal (ideal) element $\sigma$, \textit{i.e.}, $\rho\to\sigma$. Simultaneously, the entropy decreases, such that $\mu \rho \to \mu \sigma$ and $\mu$ is said to be monotone.

In classical physics, which assumes a neo-realist viewpoint, we typically can infer knowledge about a system given some minimum amount of information. For example, suppose that the information about the state of a system is fully encoded in the number $\pi$. Suppose via some process that we have been given information about the state in the form of the decimal $3.14\pm0.01$. This is clearly not enough information for us to say with certainty that the state of the system is given by $\pi$. For example, the number $3.146\ldots$ falls within our range of uncertainty. This number is the square root of $9.9$ and is algebraic, whereas $\pi$ is transcendental. While perfect certainty in this example is impossible, since both $\pi$ and $\sqrt{9.9}$ are irrational, we can at least establish a limit, such that, at some point, we can be fairly certain that the state is $\pi$. In other words, in classical systems, there may a threshold $\rho_{\textrm{min}}$, such that if $\rho_{\textrm{min}}\le \rho$, $\sigma$ may be predicted with near certainty, \textit{i.e.}, once $\rho$ passes some threshold, we may say with confidence that $\rho\approx \sigma$. A system whose future states may be predicted with certainty based on complete knowledge of its prior states may be said to be physically deterministic. Note that this definition of determinism inherently assumes that measurements do not disturb the system~\cite{endnote5}.
\begin{Condition}[physical determinism]
Let $\mathcal{I}\equiv \sigma$ be the maximal (ideal) element on a domain of physical states $\rho_n \in (\Sigma,\sqsubseteq )$, where for some value $n=N$, $\rho_n = \sigma$. For a sequence of measurements $\mu\rho_1,\mu\rho_2,\ldots,\mu\rho_N$, if $\rho_{\textrm{min}}\le\rho_N$ and $N$ is finite, then if $\rho = \mathcal{I}$ and $\mu\rho =\mu\mathcal{I} = 0$, the system is said to be physically deterministic. \label{cond1}
\end{Condition}
A hypothetically omniscient being who happens to be in possession of $\rho_{\textrm{min}}\le \rho_{N}$ (\textit{i.e.}, who possesses enough information about a system to fully predict its future states) is known as Laplace's demon. A system for which $N$ is not finite, but that otherwise satisfies all other aspects of the condition, may be said to be approximately deterministic. In the example in which the state of a system is given by $\pi$, there is clearly a point at which the system becomes approximately deterministic (though, to some extent, this point is arbitrary). It is clear from this that physical determinism inherently assumes a neo-realist viewpoint. This is not without its problems, since it assumes that, if the universe is a valid system, as we obtain more and more information about it, the entropy should decrease. Of course, it is well known that the exact opposite is actually happening. The explanation for this lies in quantum contextuality.

\section{Contextuality}\label{sec2}

Consider three states, $\rho,\sigma,\tau \in \Sigma$, and suppose that $\rho\preceq \sigma$. Furthermore suppose that $\sigma \sqsubseteq \tau$. This means that $\rho$ carries {essential} information about $\sigma$ and $\sigma$ carries some (not necessarily essential) information about~$\tau$. In order for us to conclude from this that $\rho\preceq \tau$, we would need to know that the statement $\sigma \sqsubseteq \tau$ is being made in the same {context} as $\rho \preceq \sigma$. Coecke and Martin define context in the form of the following proposition~\cite{Coecke:2011uq}.
\begin{Proposition}[context]
For all states $\rho,\sigma,\tau\in\Sigma$, if $\rho\preceq\sigma\sqsubseteq\tau$ and $\twoheaduparrow\tau\ne\{0\}$, then $\rho\preceq\tau$.
\label{prop1}
\end{Proposition} 
I refer the interested reader to~\cite{Coecke:2011uq} for the proof of this proposition. Intuitively what this says is that if a unique information order can be established for $\tau$ that includes $\rho$ and $\sigma$, then $\rho$ and $\tau$ have the same ``context'', which means the former carries essential information about the latter. Considering the example of the ball in the boxes, since the ball is assumed to be in one and only one box, the states of all of the other boxes, whether or not they are opened, contain {essential} information about the ball: {it is not in them}. That is the essence of context. As it happens, the results of classical measurements are elements of {continuous domains} and approximation on continuous domains is {context independent}~\cite{Coecke:2011uq}. Note that this is a statement about {domains} in the mathematical sense set out above and is {not} a statement concerning actual physical experiments. Recall that measurements here simply refer to a rank ordering of states. What this means is that for classical measurements, it is {automatically true} that if $\rho\preceq \sigma$ and $\sigma \sqsubseteq \tau$, then $\rho\preceq \tau$ (for $\rho,\sigma,\tau\in\Sigma$). This is at the root of the fact that there is a {unique} order on classical states, as I briefly mentioned in Section~\ref{sec1}. Again, this formalizes the fact that classical states naturally have a neo-realist interpretation.

Quantum states are not necessarily representable by measurements on continuous domains and, so, are not necessarily context independent. Recall that on the most fundamental level, we are actually working with topoi and $\rho$ is a mapping from a state object to a quantity-value object. In topos theory, the state object $\mathcal{Q}$ has no points (microstates), and as briefly mentioned above, D\"{o}ring and Isham show that this is equivalent to the Kochen--Specker (KS) theorem~\cite{Doring:2011fk}. The KS theorem essentially points out a conflict between two neo-realist assumptions: (i) that measurable quantities have definite values at all times and (ii) that the values of those measurables are intrinsic and independent of the measurement process used to obtain them~\cite{Kochen:1967aa}. In the language we developed in Section~\ref{sec1}, Assumption (i) says that a definite state of a sub-system $\rho : \mathcal{Q}\to\mathbbm{R}$ exists at all times, while Assumption (ii) says that the value corresponding to $\mathbbm{R}$ is {unique}. For quantum systems, we are forced to abandon one or the other of the two assumptions.

The KS theorem thus establishes the notion of {context} as fundamental to quantum measurements: the state of the sub-system and/or the real-valued quantity that is associated with that state, is dependent on the details of the measurement on that state. The notion of context set forth in Proposition~\ref{prop1} formalizes this idea in an order-theoretic way: for any set of states $\rho,\sigma,\tau\in\Sigma$, if $\rho\preceq\sigma\sqsubseteq\tau$, we can only conclude that $\rho\preceq\tau$ if $\twoheaduparrow\tau\notin\{0\}$. Recall that the relation $\sqsubseteq$ pertains to {knowledge} and is thus agent-dependent, whereas the relation $\preceq$ pertains to processes and is entirely independent of any agent. In other words, the relation $\preceq$ can only apply to a pair of quantum states if those states exist in the same context. Or, to put it another way, {there is no unique order on quantum states}. Let us consider two examples.

First, let us return to the entirely classical example of the ball in one of three boxes. We established that if the ball was in box $C$ and the boxes were opened in the order $A$, then $B$, then $C$, the information order would be $\rho_A\sqsubseteq\rho_B\preceq\rho_C$. Conversely, if we swapped the order in which we opened boxes $A$ and $B$, then the information order would be $\rho_B\sqsubseteq\rho_A\preceq\rho_C$. Since this is a classical system and the only information of interest is whether or not the ball is in a particular box, it should be clear that these two cases are essentially equivalent, \textit{i.e.}, $\rho_A = \rho_B$. As such:
\begin{equation}
\rho_A\sqsubseteq\rho_B\preceq\rho_C \quad \Leftrightarrow \quad \rho_B\sqsubseteq\rho_A\preceq\rho_C
\end{equation}
and we may substitute $\preceq$ for $\sqsubseteq$ in both cases. This is the essence of a classical system: it is completely independent of context and thus independent of any agent.

\begin{figure}
\begin{tikzpicture}
\node at (-1.5,-0.1)[left] {$\ket{\psi}$};
\draw[ultra thick,gray!30!yellow!60!orange,-latex'] (-1.5,-0.1) -- (-0.9,-0.1);
\draw[ultra thick,gray!30!yellow!60!orange] (-1,-0.1) -- (-0.6,-0.1);
\node at (0,0) {
 \begin{tikzpicture}
 \shade[left color=gray!50!white,right color=blue!25!white!25!gray] (0,1) rectangle (1,0);
 \shade[top color=gray!50!white,bottom color=blue!25!white!25!gray](0,1) -- (0.25,1.25) -- (1.25,1.25) -- (1,1) -- cycle;
 \shade[top color=gray!50!white,bottom color=blue!25!white!25!gray] (1.25,1.25) -- (1.25,0.25) -- (1,0) -- (1,1) -- cycle;
 \fill[black!50!gray] (1.125,0.85) ellipse (0.0525cm and 0.15cm);
 \fill[black!50!gray] (1.125,0.4) ellipse (0.0525cm and 0.15cm);
 \fill[red] (0.6,1.125) arc (0:180:0.15cm and 0.15cm);
 \fill[blue] (1,1.125) arc (0:180:0.15cm and 0.15cm);
 \end{tikzpicture}};
\node at (0,-1) {$\mathbf{S}_{A}$};
\draw[ultra thick,gray!30!yellow!60!orange,-latex'] (0.475,0.2) -- (1.1,0.2);
\draw[ultra thick,gray!30!yellow!60!orange] (1,0.2) -- (1.5,0.2);
\node at (2,0.25) {
 \begin{tikzpicture}
 \shade[left color=gray!50!white,right color=blue!25!white!25!gray] (0,1) rectangle (1,0);
 \shade[top color=gray!50!white,bottom color=blue!25!white!25!gray](0,1) -- (0.25,1.25) -- (1.25,1.25) -- (1,1) -- cycle;
 \shade[top color=gray!50!white,bottom color=blue!25!white!25!gray] (1.25,1.25) -- (1.25,0.25) -- (1,0) -- (1,1) -- cycle;
 \fill[black!50!gray] (1.125,0.85) ellipse (0.0525cm and 0.15cm);
 \fill[black!50!gray] (1.125,0.4) ellipse (0.0525cm and 0.15cm);
 \fill[red] (0.6,1.125) arc (0:180:0.15cm and 0.15cm);
 \fill[blue] (1,1.125) arc (0:180:0.15cm and 0.15cm);
 \end{tikzpicture}};
\node at (2,-1) {$\mathbf{S}_{B}$};
\draw[ultra thick,gray!30!yellow!60!orange,-latex'] (2.475,0) -- (3.1,0);
\draw[ultra thick,gray!30!yellow!60!orange] (3,0) -- (3.5,0);
\node at (4,0.1) {
 \begin{tikzpicture}
 \shade[left color=gray!50!white,right color=blue!25!white!25!gray] (0,1) rectangle (1,0);
 \shade[top color=gray!50!white,bottom color=blue!25!white!25!gray](0,1) -- (0.25,1.25) -- (1.25,1.25) -- (1,1) -- cycle;
 \shade[top color=gray!50!white,bottom color=blue!25!white!25!gray] (1.25,1.25) -- (1.25,0.25) -- (1,0) -- (1,1) -- cycle;
 \fill[black!50!gray] (1.125,0.85) ellipse (0.0525cm and 0.15cm);
 \fill[black!50!gray] (1.125,0.4) ellipse (0.0525cm and 0.15cm);
 \fill[red] (0.6,1.125) arc (0:180:0.15cm and 0.15cm);
 \fill[blue] (1,1.125) arc (0:180:0.15cm and 0.15cm);
 \end{tikzpicture}};
\node at (4,-1) {$\mathbf{S}_{C}$};
\draw[ultra thick,gray!30!yellow!60!orange,-latex'] (4.475,0.31) -- (5.25,0.31);
\draw[ultra thick,gray!30!yellow!60!orange,-latex'] (4.475,-0.14) -- (5.25,-0.14);
\node at (5.25,0.31)[right] {$+$};
\node at (5.25,-0.14)[right] {$-$};
\node at (5.75,0.085)[right] {\Large{\}} \normalsize{?}};
\end{tikzpicture}
\caption{\label{fig1} Each box represents a measurement of the spin for a spin-$\frac{1}{2}$ particle along some axis with the top output indicating that the state is {aligned} ($+$) with the measurement axis and the bottom output indicating that the state is {anti-aligned} ($-$) with the measurement axis. Red and blue lights on the top simply indicate to the experimenter which of the two results is obtained (e.g., red might indicate aligned and blue might indicate anti-aligned).}
\end{figure}
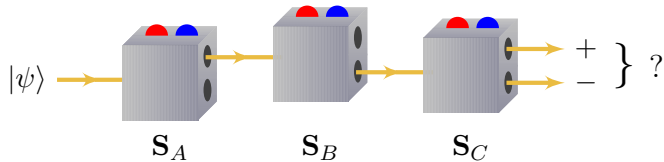

Now, consider a sequence of spin-$\frac{1}{2}$ measurements on a certain quantum system, as shown in Figure~\ref{fig1}. Due to the nature of quantum mechanical spin, we generally assume that it is an intrinsic property, such that it has a definite value along a given axis if measured along that axis. In other words, a neo-realist interpretation would assume that, if the spin is measured along, say, axis $A$ (corresponding to basis $\mathsf{a}$), and was found to be aligned with that axis, then regardless of any intermediate measurements along other axes, any subsequent measurement along $A$ must necessarily show the spin to be aligned with that axis. Quantum mechanics, however, tells us that the probabilities associated with the two possible outcomes to a measurement along axis $C$ (basis $\mathsf{c}$) in the figure solely depend on the relative alignment of axes $B$ and $C$ and the state as it enters the device measuring along axis $C$. For example, if, as in the figure, the state exiting the middle device measuring axis $B$ in basis $\mathsf{b}$ is $\ket{\mathsf{b}-}$, then the probabilities for the outcomes from the third device are $\rm{Pr}(\mathsf{c}+) = \sin^{2}\frac{1}{2}\theta_{BC}$ and $\rm{Pr}(\mathsf{c}-) = \cos^{2}\frac{1}{2}\theta_{BC}$, where $\theta_{BC}$ is the angle between the $B$ and $C$ axes~\cite{Sakurai:1994qv}. If, for instance, $\theta_{BC}=\frac{\pi}{2}$, then $\rm{Pr}(\mathsf{c}+)=\rm{Pr}(\mathsf{c}-)=0.5$, meaning that the state as measured along  axis $C$ could equally well be aligned or anti-aligned with that axis. This is independent of the outcome of any previous measurement. That means that if $A$ and $C$ represent the same axis, it is possible for the state to be measured to be $\ket{\mathsf{a}+}$ initially, but then later to be found to be $\ket{\mathsf{a}-}$. In the classical example with the ball in one of three boxes, this would be the equivalent of opening box $A$ and finding the ball, then opening box $B$ (finding nothing) and, finally, opening box $C$ and finding the ball again.

From these examples, it should be clear that no unique ordering relation exists for quantum states. Recall that the statement $\rho\preceq\sigma$ is interpreted as saying that $\rho$ contains essential information about $\sigma$. Therefore, in the classical example, if the ball is in box $C$, it clearly is {not} in any of the other boxes. Thus, the other boxes contain essential information about the location of the ball: it is not there. In the quantum example, even though the prior state does affect the probabilities, it does not {guarantee} anything. It merely establishes whether an information order exists or not. 
For example, consider just the first two spin measurement devices in Figure~\ref{fig1}, and as in the figure, assume that the measurement along $A$ finds the state aligned with that axis. The probabilities associated with a measurement along axis $B$ are $\rm{Pr}(\mathsf{b}+)=\cos^2\frac{1}{2}\theta_{AB}$ and $\rm{Pr}(\mathsf{b}-)=\sin^2\frac{1}{2}\theta_{AB}$, respectively. Let us consider three cases~\cite{endnote6}:
$$
\begin{array}{rcc}
\rm{(i):} & & \theta_{\mathsf{ab}}=\frac{\pi}{2} \\
\rm{(ii):} & & 0<\theta_{\mathsf{ab}}<\frac{\pi}{2} \\
\rm{(iii):} & & \theta_{\mathsf{ab}}=0.
\end{array}
$$
In Case (i), the probabilities for the outcome of a measurement along axis $B$ will be $\rm{Pr}(\mathsf{b}+)=\rm{Pr}(\mathsf{b}-)=\frac{1}{2}$. This is equivalent to a completely random outcome meaning that we have no information whatsoever about $\rho_B$ prior to making the measurement along $B$. Order theoretically, prior to the measurement, $\rho_B$ is a completely mixed state as in~\eqref{eq:mixed} and is said to be a {least element} on the domain. What this means is that {no information order can be established for $\rho_A$ and $\rho_B$}; neither $\sqsubseteq$ nor $\preceq$ apply, since knowledge of the outcome of the measurement along $A$ does nothing to improve our chances of predicting the outcome of a measurement along $B$.

Now consider Case (ii). In this case, the probabilities are not $\frac{1}{2}$, and so, knowledge of the outcome of the measurement along $A$ {does} improve our chances of correctly predicting the outcome of a measurement along $B$, but it still does not guarantee a specific result. In this way, we may write $\rho_A\sqsubseteq\rho_B$, since this partial knowledge does allow us to establish some kind of order on the states. Likewise, for Case (iii) if $\theta_{AB}=0$, the outcome of the measurement along $B$ is {guaranteed} to be exactly the same as it was along $A$. As such, we may write $\rho_A\preceq\rho_B$.

In order to generalize this, it is necessary to introduce the order-theoretic notion of a {basis}. Consider some subset of states $\mathsf{m}$ on a domain ($\Sigma,\sqsubseteq$), such that $\mathsf{m}\subseteq\Sigma$. The subset $\mathsf{m}$ is said to be a {basis} when $\mathsf{m}\;\cap\twoheaddownarrow\rho$ is directed with supremum $\bigsqcup\rho$ for each $\rho\in\Sigma$. Recall that a dcpo is a poset in which every directed set has a supremum. Notice that this definition inherently includes the relation $\preceq$ (via $\twoheaddownarrow\rho$). As such, it codifies the notion that a basis is any subset for which the neo-realist assumption holds {within that subset}. In order to see why the subset must be directed with a supremum, consider just two boxes, one of which contains a ball of some indeterminate color. The state representing the box that contains the ball is the supremum, since it contains more information about the color of the ball than the box that does not contain the ball. It is necessarily directed, because, regardless of the order in which the boxes are opened, $\rho_{\rm{no}\;\rm{ball}}\preceq\rho_{\rm{ball}}$.

Now, consider two quantum states, $\rho$ and $\sigma$. If they are measured in the same basis, \textit{i.e}., if $\rho,\sigma\in\mathsf{m}\subseteq\Sigma$, then we can establish relationships, such as $\rho\preceq\sigma$. If they are measured in different bases, the establishment of any kind of information order is dependent on how ``close'' they are. We might be tempted to use the term ``orthogonal'' here to describe two bases for which no information order can be established, based on our example with the spin measurements. This can be a bit confusing, since each of these bases are individually said to be orthogonal, since measurements on different elements within a given basis should satisfy~\eqref{eq:ortho}. In fact, it is the orthogonality of various information orders that {defines} a basis in the first place. Consider the classical example of the three boxes, $A$, $B$ and $C$, with a ball in one of them. Without knowledge of the ball's location (\textit{i.e}., before measurement), there are three possible information orders:
$$
\rho_A, \rho_B\sqsubseteq\rho_C, \quad\quad \rho_A, \rho_C\sqsubseteq\rho_B, \quad\quad \rho_B, \rho_C\sqsubseteq\rho_A
$$
given by $\uparrow\rho_C$, $\uparrow\rho_B$ and $\uparrow\rho_A$, respectively, depending on where the ball is located. Since only one may be correct in a neo-realist interpretation, a measurement on the intersection of any two must satisfy~\eqref{eq:ortho}. Thus, we can think of measurements on these boxes (which entails opening them) as representing a method for identifying an orthogonal basis for measurements where those measurements are aimed at determining the location and color of the ball. In other words, the set $\uparrow\rho_A,\uparrow\rho_B,\uparrow\rho_C =: \mathsf{m}\subseteq\Sigma$ defines the basis. Classically, of course, this is somewhat irrelevant, but it serves to illustrate the definition.

Clearly, then, if a measurement on states at the intersection of any two information orders is anything {other} than an element of the null set, the two information orders must necessarily be defined on different bases that share some information, \textit{i.e}., for $\rho\in\mathsf{m}\subseteq\Sigma$ and $\sigma\in\mathsf{n}\subseteq\Sigma$:
\begin{equation}
\mu(\uparrow\rho\;\cap\uparrow\sigma)\notin\{0\} \; \Rightarrow \; \mathsf{m}\ne\mathsf{n}.
\label{eq:caseii}
\end{equation}
This adequately generalizes Case (ii) above. Distinguishing between Case (i) and Case (iii) is a bit more problematic, since the result $\mu(\uparrow\rho\;\cap\uparrow\sigma)\in\{0\}$ does not automatically (in a mathematical sense) tell us whether $\rho$ and $\sigma$ belong to the same basis or not (recall that orthogonality is defined here by an information order).

Consider the set $\mathcal{M}(\uparrow\rho\;\cap\uparrow\sigma)$ of all measurements on states at the intersection of any two information orders where $\mu\in\mathcal{M}$. Recall that $\mu$ is defined as a function on a domain that assigns to each informative object a number that measures the amount of ``partiality'' such that continued measurements lead to {decreasing} partiality (see~\eqref{eqinfo}). The {greatest} amount of partiality, then, corresponds to the greatest {lack} of knowledge (which is why $\mu$ is often most conveniently expressed as entropy). The greatest lack of knowledge regarding any two information orders is associated with complete randomness. In the example given in Figure~\ref{fig1}, this corresponds to Case (i), where we may write the state in one basis in terms of a different basis' {least element}. For example, suppose that $\mathsf{a}=\mathsf{z}$ and $\mathsf{b}=\mathsf{x}$. The state of the system exiting the second spin measurement device in Figure~\ref{fig1} can be written in terms of the first:
\begin{equation}
\ket{\mathsf{x}-}=\frac{1}{\sqrt{2}}\ket{\mathsf{z}+}-\frac{1}{\sqrt{2}}\ket{\mathsf{z}-}.
\label{eq:bases}
\end{equation}
Two points should be clear from this. First, this demonstrates that the set $\mathcal{M}$ is partially ordered. Second, it allows us to clearly define a supremum for $\mathcal{M}$ as representing the case in which knowledge of a unique basis corresponds to a least element on one of the bases (\textit{i.e}., as in the example given by~\eqref{eq:bases}). Thus, we may say that for $\rho\in\mathsf{m}\subseteq\Sigma$ and $\sigma\in\mathsf{n}\subseteq\Sigma$:
\begin{equation}
\mu(\uparrow\rho\;\cap\uparrow\sigma)=\bigsqcup\mathcal{M}(\uparrow\rho\;\cap\uparrow\sigma) \; \Rightarrow \; \mathsf{m}\perp\mathsf{n}
\label{eq:casei}
\end{equation}
where I use the symbol ``$\perp$'' intentionally to tie this to the concept of a least element and a completely mixed state (e.g., Equation~\eqref{eq:mixed}). This, then, adequately generalizes Case (i) above. Thus Case (iii) is simply the definition of orthogonality given by~\eqref{eq:ortho}, \textit{i.e}.,
\begin{equation}
\mu(\uparrow\rho\;\cap\uparrow\sigma)\in\{0\}\;\Rightarrow\;\mathsf{m}=\mathsf{n}.
\label{eq:orthocontext}
\end{equation}
Intuitively, this says that within any given basis, there is a single, unique information order, and thus, neo-realism holds (within that basis; see Section~\ref{sec3}). Any difference in basis necessarily eliminates the uniqueness of the ordering relation, and a neo-realist interpretation is no longer tenable under these conditions. It is the dependency of quantum systems on a measurement basis (\textit{vis-\`{a}-vis} projective measurements) that is at the heart of this behavior. Classical systems do not suffer from this complication and, thus, are context independent, as discussed before.

This idea may now be related to Proposition~\ref{prop1} as a means of {quantifying} (quantum) contextuality, since some measurement $\mu(\uparrow\rho\;\cap\uparrow\sigma)$ is a number that necessarily lies on the interval $0\le\mu(\uparrow\rho\;\cap\uparrow\sigma)\le\bigsqcup\mathcal{M}(\uparrow\rho\;\cap\uparrow\sigma)$. A specific measurement context can then be defined as follows, and the value of $\mu(\uparrow\rho\;\cap\uparrow\sigma)$ is akin to a ``distance'' measure telling us how ``far'' one measurement context is from~another.
\begin{Definition}[measurement context] For any three quantum states $\rho\in\mathsf{j}\subseteq\Sigma$, $\sigma\in\mathsf{k}\subseteq\Sigma$ and \mbox{$\tau\in\mathsf{l}\subseteq\Sigma$}, where $\mathsf{j}$, $\mathsf{k}$ and $\mathsf{l}$ are bases, if $\mu(\uparrow\rho\;\cap\uparrow\sigma)\in\{0\}$ and $\mu(\uparrow\sigma\;\cap\uparrow\tau)\in\{0\}$, then it follows that $\mu(\uparrow\rho\;\cap\uparrow\tau)\in\{0\}$ and $\rho$, $\sigma$ and $\tau$ are said to have an identical context, \textit{i.e}., $\rho\preceq\sigma\preceq\tau$.
\end{Definition}
Note that the fact that this definition necessarily ensures that a unique information order exists, e.g., $\twoheaduparrow\tau\notin\{0\}$. Any values of $\mu$ other than elements of the null set imply that, at best, the states only possess a {partial context}. Measurements on states that represent a supremum on the set of all possible measurements imply that those states share {no context}. Thus, we have established an order-theoretic method for quantifying contextuality. 

\section{Context, Determinism and Entropy}\label{sec3}

What does it mean to say that an information order may be defined {within} a basis? In Figure~\ref{fig1}, a single particle always exits one and only one of the output channels of any given spin measurement device. Subsequent measurements of the same particle in that basis (\textit{i.e}., without ever changing the basis) yield the same result. In other words, as long as a basis does not change, it behaves much as the classical example of a single ball that is inside one of multiple boxes. As such, it possesses a unique information order (just as the ball and box example does), and thus, neo-realism holds, even though only a single measurement is usually required to produce the outcome. For example, in Figure~\ref{fig1}, the information order $\rho_{\mathsf{b}+}\preceq\rho_{\mathsf{b}-}$ remains unique as long as the measurement basis is not changed, even though only a single measurement is required to establish the order. This is simply due to the nature of quantum measurement devices. The same would be true, for example, in the classical case involving the boxes and the ball if all of the boxes could be opened and viewed simultaneously. Any restriction to this is practical and not theoretical.

This concept, then, is easily generalized to an $n$-dimensional basis $\mathsf{m}_n$: though only a single measurement may be required in order to determine the state, this measurement nevertheless establishes a unique information order. As such, following the prescription given by~\eqref{eq:entcond}, $\mu$ is monotone, so long as the basis does not change. It is thus trivially true that Condition~\ref{cond1} holds for any sequence of measurements on quantum states in which the basis does not change and the basis itself is finite-dimensional. Equation~\eqref{eq:entcond} tells us that this corresponds to a decrease in entropy.

Condition~\ref{cond1} does not hold, however, if the basis changes. This is because, as the above examples clearly demonstrate, each set of basis states {has its own maximal element}. In other words, inherent in the order theoretic definition of physical determinism given by Condition~\ref{cond1} is the notion of the uniqueness of a maximal element on the complete set of states for some physical system. A change of basis in a quantum system introduces a new maximal element associated with the new basis. Thus, the complete set of states for a quantum system does not possess a single unique maximal element, but, rather, possesses many. Therefore, a full characterization of the complete set of states for a quantum system is not physically deterministic. In addition, for a sequence of measurements on this complete set of states, the entropy will not decrease. I formalize this with the following remark.
\begin{Remark}
For some complete set of quantum states \underline{$\Sigma$} measured on a complete set of finite-dimensional bases, no single, unique maximal element $\mathcal{I}$ exists.
\label{rem1}
\end{Remark}
This result is very closely related to the topos-theoretic statement of the Kochen--Specker (KS) theorem, as given in~\cite{Doring:2011fk}. The details in~\cite{Doring:2011fk} are quite involved, but some general remarks are in order.

The topos-theoretic statement of the KS theorem as given in~\cite{Doring:2011fk} employs the concept of a {spectral presheaf}. While the details are beyond the scope of this essay, suffice it to say that a spectral presheaf is a particular type of category-theoretic mapping called a {contravariant functor}. For our purposes, it is essentially representative of the complete set of states on a system, and we will label it \underline{$\Sigma$} to distinguish it from $\Sigma$. Very roughly, a {global element} on \underline{$\Sigma$} is a function that essentially assigns a unique number to each state on \underline{$\Sigma$}. In other words, it guarantees that there should be one and only one maximal element. As stated in~\cite{Doring:2011fk}, the KS theorem only applies to systems described by a finite-dimensional Hilbert space with $\textrm{dim}\;\mathcal{H}>2$. What it says is that \underline{$\Sigma$} {has} no global element.

In the notion of contextuality that I have developed here, there is no restriction on the dimensionality of the Hilbert space, \textit{per se}. There {is}, however, a restriction to finite-dimensional bases. This provokes the following remark. In both statements of KS contextuality (\textit{i.e}., here and in~\cite{Doring:2011fk}), complications with the infinite-dimensional basis states arise from the fact that they can have continuous parts. As Coecke and Martin show, continuous bases are context independent~\cite{Coecke:2011uq}. Thus, Remark~\ref{rem1} is analogous to the topos-theoretic statement of the KS theorem.

One of the key points related to the lack of a unique maximal element on sets of quantum states, as pointed out earlier, is the fact that if measurements are made on a complete set of bases, the overall entropy of the system will {not} decrease. Each change of basis essentially ``resets'' the system and, thus, the entropy. Consider a complete set of $N$ measurement bases on a system of states $\Sigma$. If the system behaves classically, then~\eqref{eq:entcond} holds and a sequence of measurements will result in decreasing partiality, \textit{i.e.}, increasing knowledge about the system. This corresponds to the existence of a single, unique maximal element. It must necessarily be true, then, that the existence of multiple maximal elements would prevent the entropy from decreasing. We might imagine, though, a system for which there are $N$ maximal elements corresponding to sub-systems, such that the entropy could decrease for any given sub-system individually. This, however, would require neo-realism to hold for a given sub-system regardless of whether the measurements on that sub-system are interrupted by measurements on a different sub-system.

Quantum systems, of course, do not behave in this manner. In the example given in Figure~\ref{fig1}, there is no guarantee that the outcome of the third device will match that of the first device, even if they measure along the same axis. As such, it is possible that re-measuring in a given basis may result in a different maximal element {for that basis}. In other words, in quantum systems, it is possible for the maximal element of a given basis to change. This means that even if the complete set of bases is finite, the number of maximal elements may be {infinite}. This may at first appear paradoxical, but the paradox is resolved if we consider that the state as measured by the first device is not the same state as that measured by the third device, regardless of the outcome. This is because the object being measured fundamentally possesses a world line in spacetime. In the example given in Figure~\ref{fig1}, the object is a localized qubit. However, as pointed out in~\cite{Palmer:2012aa,Palmer:2013aa}, in a strict sense, the ``location'' of a qubit on a world line is fundamentally a part of its state, \textit{i.e}., a localized qubit is really best understood as a {sequence} of quantum states associated with points on a world line. In other words, {quantum states are constantly changing}, since they are associated with objects that possess world lines. If the world line is infinite, then, regardless of the number of possible measurement bases, the number of maximal elements must be infinite, since one exists for each possible measurement. This, in fact, is the very {essence} of contextuality: neo-realism fails {spectacularly}. In this case, the overall entropy of the complete system will tend to {increase}. This warrants an additional remark. 
\begin{Remark}
For a sequence of measurements on a complete set of finite-dimensional bases for some complete set of quantum states $\rho_n\in$ \underline{$\Sigma_n$}, the entropy must be greater than or equal to zero, \textit{i.e}., $\mu\rho_n\ge0$ (recall that complete knowledge of a system corresponds to $\mu\rho=0$).
\label{rem2}
\end{Remark}
Remark~\ref{rem2} bears a striking resemblance to the second law of thermodynamics, sometimes called the ``law of increase of entropy.'' Indeed, the quantum mechanical origin of this law is not a new suggestion. In the 1958 English translation of their volume on statistical physics, Landau and Lifshitz write that ``[i]t is more natural to think of the origin of the law of increase of entropy in its general formulation \ldots as being bound up with the quantum mechanical effects''. They continue:
\begin{quote}
[I]f two [quantum mechanical] processes of interaction take place consecutively with a given quantum object (let us call them $A$ and $B$) then the assertion that the probability of some result of process $B$ is determined by the results of process $A$ can be true only if process $A$ takes place before process $B$.  \cite{Landau:1958aa} (p.31).
\end{quote}
This statement is equivalent to the order-theoretic statement ``$\rho_A\sqsubseteq\rho_B$ for $\rho_A,\rho_B\in$ \underline{$\Sigma$} is not a unique information order'' (where \underline{$\Sigma$} is understood to represent a quantum mechanical system). As I have argued, this lack of a unique ordering relation on quantum states is an order-theoretic manifestation of the phenomenon of quantum contextuality. Thus, it would appear as if quantum contextuality is at the root of the second law of thermodynamics. However, note that I have also shown that the lack of a unique ordering relation arises from the presence of an agent. This would seem to suggest, then, that the second law itself is somehow agent-dependent~\cite{endnote7}. 

\section{Summary and Concluding Remarks}\label{sec4}

In this essay, I have developed order-theoretic notions of determinism and contextuality on domains and topoi, in the process developing an order-theoretic quantification of contextuality that is compatible with the sense of the term embodied in the Kochen--Specker theorem. The order-theoretic view has allowed me to show that, while a unique ordering relation exists for classical states, no such unique relation exists for quantum states. I have argued that this lack of a unique ordering relation necessarily appears with the introduction of an agent. As such, quantum states do not allow for a neo-realist interpretation. This fact is a result of the contextual nature of quantum states. Thus, contextuality (at least in the sense given by the Kochen--Specker theorem) is deeply connected to the concept of a measuring agent. Contextuality also assures us that no sequence of measurements on quantum states can lead to the complete characterization of a quantum system in the same sense that such a sequence of measurements on classical states could completely characterize a classical system. In fact, the entropy associated with such a sequence of measurements on a quantum system will necessarily never decrease. Incidentally, this is perfectly consistent with the notion of conditional quantum entropy, which can be negative ({\em cf.}~\cite{Cerf:1997aa,Cerf:1999aa}). Entropy, as the term is applied here, simply refers to the entropy associated with measurements on the system that establish an information order. Nevertheless, this non-decrease in entropy for measurements on quantum systems bears a striking resemblance to the second law of thermodynamics when applied to sequential measurements. Indeed it essentially formalizes a suggestion made by Landau and Lifshitz in 1958. This does seem to suggest that the second law is an agent-dependent phenomenon. Additionally, it suggests some relation between contextuality and thermodynamics.

This essay suggests at least two pieces of additional work. First, Remark~\ref{rem1} should be put on more solid ground by stating it as a proposition, lemma or theorem supported by a formal proof. Second, a deeper relation between Remark~\ref{rem2} and the second law of thermodynamics should be found by formalizing the latter in order-theoretic terms on domains of generalized states. This would necessarily involve additional work related to coarse-graining and, potentially, could involve extensions to generalized probability theories. It is interesting to note in passing that coarse-graining is inherent in the derivation of the Kochen--Specker theorem in terms of spectral presheafs given by D\"{o}ring and Isham. Given the close analogy to between Remark~\ref{rem1} and their statement of the KS theorem, it stands to reason that additional work on coarse-graining in relation to Remark~\ref{rem2} should yield a deeper connection and should solidify any relation between contextuality and thermodynamics.

\acknowledgements{Acknowledgments}

I wish to thank Bob Coecke for introducing me to categories and domains and for sending me a copy of \textit{New Structures for Physics}, by which I was inspired. I also wish to thank two anonymous referees for very helpful comments that aided in making my arguments more succinct. In particular, I would like to thank one referee for introducing me to the work of Acz\'{e}l, Forte and Ng, which has yielded additional insights. Finally, I acknowledge financial support from FQXi.

\end{document}